\documentclass[journal]{IEEEtran}

\usepackage{graphicx}
\usepackage{dcolumn}
\usepackage{bm}
\usepackage{amsfonts}
\usepackage{amsmath}

\begin{document}

\title{High Efficiency Postprocessing for Continuous-variable Quantum Key Distribution:\\Using All Raw Keys for Parameter Estimation and Key Extraction}

\author{Xiangyu~Wang,
            Yichen~Zhang,
            Song~Yu,
        and~Hong~Guo
}

\maketitle

\begin{abstract}
High efficiency postprocessing of continuous-variable quantum key distribution system has a significant impact on the secret key rate and transmission distance of the system. Currently, the postprocessing mainly contains four steps in the following order: sifting, parameter estimation, information reconciliation, and privacy amplification. For a quantum channel with unknown prior characteristics, part of the raw keys (typically half) have to be sacrificed to estimate the channel parameters, which is used to assist in error correction (part of information reconciliation) and estimate the secret key rate. This introduces a tradeoff between the secret key rate and the accuracy of parameter estimation when considering the finite-size effect. In this paper, we propose a high efficiency postprocessing method which uses all the raw keys for both parameter estimation and key extraction. It is realized by exchanging the order of parameter estimation and information reconciliation, while the other steps remain unchanged. After the sifting step, the extra data (used for phase compensation and synchronization etc) can be used to roughly estimate the quantum channel parameters, or using the estimated results of last block, thus the error correction can be realized normally. After the success of error correction, the reconciler recovers the raw keys of the other side by a reverse mapping function. Then she uses all the raw keys of both sides for parameter estimation to estimate the quantum channel parameters and calculate the secret key rate of the system. Finally, they perform the privacy amplification step to obtain the unconditional security keys. We show that this method improves the accuracy of parameter estimation and the secret key rate of continuous-variable quantum key distribution system.
\end{abstract}


\section{Introduction}

\IEEEPARstart{Q}{uantum} key distribution (QKD)~\cite{gisin2002quantum} is a major practical application of quantum information technology, which allows two distant legitimate parties (named Alice and Bob) to extract unconditional secure keys though an insecure quantum channel which may be controlled by an eavesdropper (named Eve) and a classical authentication channel. There are mainly two kinds of QKD protocols, they respectively encode information on discrete variables~\cite{bennet1984quantum,ekert1991quantum} and continuous variables (CV)~\cite{grosshans2002continuous,weedbrook2004quantum,li2014continuous,pirandola2015high,zhang2014continuous}. CV-QKD takes the advantage of using standard telecommunication technologies~\cite{weedbrook2012gaussian,jouguet2013experimental,diamanti2015distributing,zhang2017continuous}, and its security has been proven when using Gaussian modulation coherent states. A CV-QKD protocol contains quantum states preparation, transmission, measurement, and postprocessing, where the imperfect postprocessing is one of the main bottlenecks that limits its property. The efficiency and the speed are two main targets of postprocessing. Several works have been done to improve the speed~\cite{WangHSEC,Yang2017FPGA,Wang2018PA}. In this paper, we focus on improving the efficiency of the postprocessing.

Currently, in the postprocessing part, the raw keys after quantum state measurement have to be divided into two parts. One part is used for parameter estimation, and the other part is used for key extraction. Typically, the data used for the first part is about 50\%. The major purpose of parameter estimation is to estimate the secret key rate that can be extracted from the raw keys used for key extraction. The accuracy of parameter estimation has a great influence on secret key rate and transmission distance of CV-QKD system when considering the finite-size effect~\cite{leverrier2010finite,jouguet2012analysis}.

The finite-size effect of parameter estimation has been analyzed in many CV-QKD protocols~\cite{leverrier2010finite,jouguet2012analysis} and continuous-variables measurement-device-independent QKD protocols~\cite{zhang2017finite,papanastasiou2017finite}. They all need to sacrifice part of the raw keys and are limited by the accuracy of parameter estimation. Cosmo Lupo {\it et al.}~\cite{lupo2017parameter} proposed a method that does not sacrifice raw keys by displacing the variables after detection based on an affine function in continuous-variables measurement-device-independent QKD protocols. In CV-QKD protocols, Leverrier~\cite{leverrier2015composable} proposed a method based on quantum tomography, in which all the raw keys can be used for both parameter estimation and key extraction. He introduced two additional parties to Alice and Bob, and the modes of quantum states are also divided into two parts, then distribute them to each of the additional party for both sides. Each of the additional party can estimate the covariance matrices of the other's quantum states. Although this method does not waste raw keys, he pointed out that it is rather impractical. Therefore, it is necessary to propose a practical high efficiency method in CV-QKD protocols that does not sacrifice raw keys and parameter estimation is accurate.

To solve this problem, we exchange the implementation order of parameter estimation and information reconciliation~\cite{van2004reconciliation,wang2017efficient} in the postprocessing of CV-QKD protocols, so that the whole raw keys can be used for both parameter estimation and key extraction. First, the channel characteristics are roughly estimated by using the extra data used for phase compensation and synchronization etc, or using the parameter estimation results of the previous block. For a stable CV-QKD system, the channel characteristics change very slowly. Therefore, the results of the latter method maybe more accurate because the amount of data used for parameter estimation is far larger than the former method. Then they choose a suitable error correction code according to the parameter estimation results. Next, they correct the errors between them to share corrected keys. Finally, the error correction party can recover the other party's raw keys if the error correction successes. This is achieved by a reversible mapping function based on multidimensional reconciliation~\cite{leverrier2008multidimensional}. After this process, she can estimate the parameters of the quantum channel with the raw keys of both side without public communication. Because the whole raw keys are used for both parameter estimation and key extraction, the results of parameter estimation are more accurate, and the data utilization is high. We will show that the proposed method obviously improves the secret key rate and transmission distance.

\section{High efficiency postprocessing for CV-QKD}
A CV-QKD protocol contains the quantum part and the classical postprocessing part. In the quantum part of prepare-and-measure scheme, Alice prepares quantum states and distributes them to Bob though an insecure quantum channel, Bob measures them by a homodyne detector or a heterodyne detector. After the measurement, the raw keys of both parties are classical Gaussian variables. The classical postprocessing part mainly includes four steps: base sifting, parameter estimation, information reconciliation, and privacy amplification~\cite{bennett1988privacy,bennett1995generalized}.

The purposes of postprocessing are to estimate the characteristics of quantum channel, correct errors, and extract secret keys. The detailed process is described as follows: (1) Base sifting: Bob records the quadratures he measures and broadcasts them to Alice. Alice keeps the corresponding Gaussian variables. Then, they share the relevant raw keys. They keep one of the quadrature values if Bob performs a homodyne detection, while both of the quadrature values are kept if Bob performs a heterodyne detection. (2) Parameter estimation: Alice and Bob disclose part of the raw keys to estimate the characteristics of quantum channel and to obtain an upper bound on the information of eavesdropper Eve. All the characteristics of the quantum channel can be expressed by a covariance matrix of the raw keys shared by Alice and Bob. Then, Alice and Bob can calculate a secret key rate lower bound that can be extracted from the rest part of the raw keys. (3) Information reconciliation: In our scheme, this step is divided to multidimensional reconciliation and error correction. Alice and Bob rotate their raw keys to build a binary input additional white Gaussian noise channel by a mapping function. Then, they correct the errors between them by a linear block code, such as low density parity check (LDPC) code~\cite{gallager1962low,richardson2008modern}. Finally, they further check whether there are still errors between them by error verification. (4) Privacy amplification: Alice and Bob apply a universal hash function to their corrected keys and compress them according to the secret key rate obtained from the parameter estimation step. Eve has nearly zero information on the final secure keys.

For a practical CV-QKD protocol, finite-size effect has to be considered when computing the secret key rate. Suppose that $x$ and $y$ are the raw keys of Alice and Bob, and $E$ refers to the quantum state of Eve. For the asymptotic regime, the secret key rate is given by
\begin{equation}
K_{asy}=I(x:y)-S(y:E),
\end{equation}
where $I(x:y)$ is the Shannon mutual information of Alice and Bob, $S(y:E)$ is the maximum Holevo information of Bob and Eve. In this paper, we use reverse reconciliation because it can beat 3dB loss limit~\cite{grosshans2003quantum}, and the transmission distance is farther than direct reconciliation. For direct reconciliation, $S(y:E)$ is replaced by $S(x:E)$. This analysis is secure against collective attacks.

In the finite-size case, the secret key rate can be expressed as~\cite{leverrier2010finite}:
\begin{equation}
K_{finite}=\frac{n}{N}[\beta I(x:y)-S_{\epsilon_{PE}}(y:E)-\Delta(n)],
\label{eq:kp}
\end{equation}
where $N$ is the total number of data exchanged by Alice and Bob, $n$ is the number of data used for key extraction, and the other $m=N-n$ data is used for parameter estimation. $\beta$ is reconciliation efficiency which means the proportion of information extracted from $I(x:y)$, it ranges from 0 to 1. $S_{\epsilon_{PE}}(y:E)$ is the upper bound of information that Eve can obtain from the information of Bob when considering the finite-size effect, where $\epsilon_{PE}$ is the failure probability of parameter estimation. This results in overestimating the ability of Eve and underestimating the secret keys rate of CV-QKD system. In most CV-QKD protocols, the raw data used for parameter estimation need to be disclosed, therefore it can not be used for key extraction, which reduces the data utilization. And parameter estimation only uses part of the raw keys, thus the results of parameter estimation are inaccurate, they only exist in confidence regions with a certain probability. The value of $\epsilon_{PE}$ can be decreased by increasing the size of raw data used for parameter estimation, while this will decrease the secret key rate. Thus, this introduces a tradeoff between the accuracy of parameter estimation and the secret key rate in the case of finite-size regime. $\Delta(n)$ is related to the security of privacy amplification. Its value is given by~\cite{leverrier2010finite}
\begin{equation}
\Delta(n)\equiv(2dim\mathcal{H}_Y+3)\sqrt{\frac{log_2(2/\overline{\epsilon})}{n}}+\frac{2}{n}log_2(1/\epsilon_{PA}),
\end{equation}
where $\mathcal{H}_Y$ is the Hilbert space corresponding to the variable of Bob, $\overline{\epsilon}$ is a smoothing parameter, and $\epsilon_{PA}$ is the failure probability of privacy amplification. Although the Hilbert space of the raw data is infinite, the final secret key rate is less than 1 bit per pulse generally. Thus, for some CV-QKD protocols, the keys are encoded on bits, and this allows users to take $dim\mathcal{H}_Y$ as 2 for parameter estimation. The parameter $\overline{\epsilon}$ is related to the speed of convergence of the smooth min-entropy of the quantum states. The parameter $\epsilon_{PA}$ is the failure probability of privacy amplification, which indicates the probability of having the same output for different input when perform the privacy amplification process by a universal hash function. Both of the two parameters can be optimized.

\begin{figure}
\includegraphics[width=8.6cm,height=12.8cm]{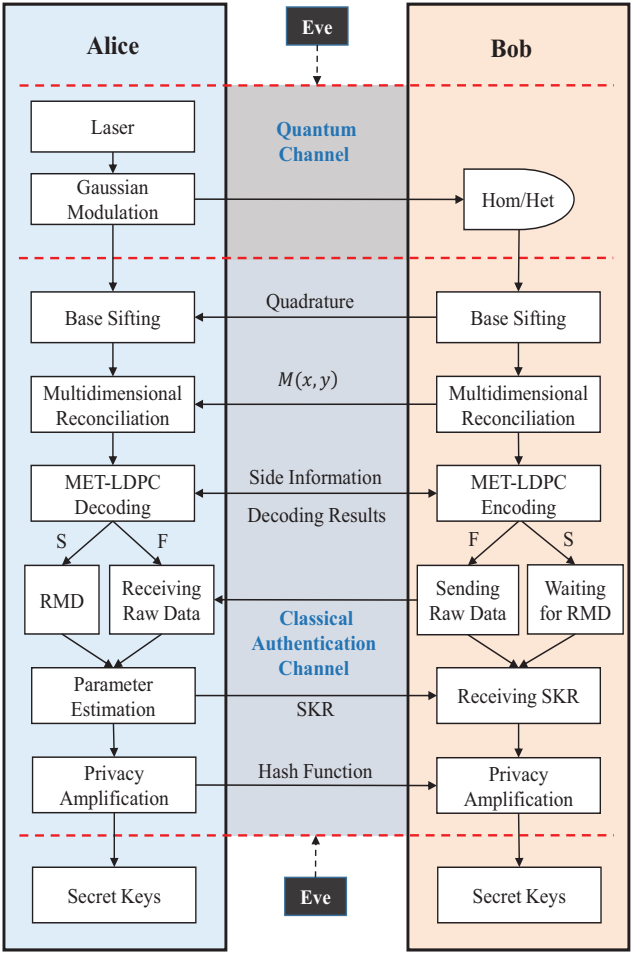}
\caption{The process of high efficiency postprocessing for CV-QKD. Hom: Homodyne detector. Het: Heterodyne detector. RMD: Reverse multidimensional reconciliation. SKR: Secret key rate. S: Success. F: Fail.}
\label{fig:process}
\end{figure}

In the following, we describe a high efficiency postprocessing method for CV-QKD systems. This not only obtains more accurate results of parameter estimation, but also improves the secret key rate. The proposed method does not waste any raw keys and improves the data utilization. The process of high efficiency postprocessing for CV-QKD is illustrated in Fig.~\ref{fig:process}.

As shown in Fig.~\ref{fig:process}, the high efficiency Gaussian modulation coherent-state CV-QKD system mainly contains two parts: the physical part and the postprocessing part. The first part includes quantum state preparation, transmission, and detection. In this part, Alice first generates light sources with a laser, and prepares Gaussian coherent states by Gaussian modulation. Then, she sends the quantum states to Bob through an untrusted quantum channel which can be controlled by a potential eavesdropper Eve. Finally, Bob uses homodyne detector or heterodyne detector to measure the quantum states. Actually, in the practical systems, due to the imperfections of quantum channel and the influence of other factors, Alice has to send Bob some extra data to complete the functions of phase compensation and synchronization etc.

The second part includes base sifting, information reconciliation, parameter estimation, and privacy amplification. This part is implemented through a classical authentication channel which means Eve can eavesdrop on the message exchanged between Alice and Bob but she can not modify them without being discovered. The first step of the postprocessing is base sifting. This step is completed according to the type of detector used to measure the quantum states. If Bob uses homodyne detector, he randomly measures one of the quadratures and records it. When this process is executed $N$ times, Bob informs Alice the quadratures he measured. Then, Alice saves the data with the same quadratures. Finally, they will share a set of related Gaussian variables with length $N$. While if Bob uses heterodyne detector, both of the quadratures are measured. Therefore, after $N$ times of quantum measurement process, Alice and Bob have a set of related Gaussian variables with length $2N$, which is 2 times of the homodyne detector protocol.

The main difference of the proposed high efficiency CV-QKD is that we exchange the execution order of information reconciliation and parameter estimation. Next, we will describe how to perform information reconciliation step without parameter estimation. This means that the whole raw keys can be used to extract secret keys. We first introduce the principle of information reconciliation. This process has two modes: direct reconciliation and reverse reconciliation~\cite{grosshans2003quantum}. The former mode uses Alice's data as raw keys, and Bob tries to recovery them. While the latter mode uses Bob's data as raw keys, and Alice tries to recovery them. Both of the modes have the same Shannon mutual information between Alice and Bob, however the information that Eve eavesdrop from Alice and Bob is different. In the direct reconciliation mode, Eve can directly eavesdrop the information of Alice from the quantum channel. While in the reverse reconciliation mode, Eve can only obtain the information from Bob indirectly. Therefore, when the channel loss is larger than 3dB, Eve may obtain more information from Alice than Bob in the direct reconciliation mode. In this paper, we mainly focus on long distance CV-QKD system. Thus, we choose the mode of reverse reconciliation.

In a low signal-to-noise ratio CV-QKD system, this process can be divided into three steps: multidimensional reconciliation, error correction, and error verification. The first step of this process is an efficient method to extract secret keys from Gaussian variables. The second step is to correct errors between Alice and Bob. The purpose of the last step is to further ensure the consistency of the keys after error correction.

Multidimensional reconciliation can construct an virtual binary input additive white Gaussian noise (BIAWGN) channel through rotating the Gaussian variables of Alice and Bob. We will describe this step in detail because that it is the key factor to achieve the proposed method. To clearly describe this step, we define the following symbols. Let $\mathbf{x}$ and $\mathbf{y}$ be the raw keys of Alice and Bob. For a linear AWGN channel, they satisfy the following relationship $\mathbf{y}=t\mathbf{x}+\mathbf{z}$ with $\mathbf{x} \sim\mathcal{N}(0,V_A)$, $\mathbf{z}\sim\mathcal{N}(0,\sigma^2)$, and $\mathbf{y}\sim\mathcal{N}(0,t^2V_A+\sigma^2)$, where $t$ is related to the loss of quantum channel, and $\mathbf{z}$ refers to the Gaussian noise. In the reverse reconciliation mode, their relationship can be rewritten as $\mathbf{x}=t'\mathbf{y}+\mathbf{z'}$ with $\mathbf{z'}\sim\mathcal{N}(0,\sigma'^2)$. The details of multidimensional reconciliation~\cite{leverrier2008multidimensional} are described as follows:

(1) Suppose that the dimension is $d$, and it is 8 in our scheme. Alice and Bob first randomly or sequentially divided their raw keys into $N/d$ ($2N/d$ for heterodyne detector) $d$-dimensional strings $\mathbf{x}^d$ and $\mathbf{y}^d$ respectively.

(2) Alice and Bob normalize their Gaussian variables, such that $\mathbf{x'}^d=\mathbf{x}^d/||\mathbf{x}^d||$, $||\mathbf{x}^d||=\sqrt{\langle\mathbf{x}^d,\mathbf{x}^d\rangle}$, and $\mathbf{y'}^d=\mathbf{y}^d/||\mathbf{y}^d||$, $||\mathbf{y}^d||=\sqrt{\langle\mathbf{y}^d,\mathbf{y}^d\rangle}$.

(3) Bob randomly chooses a string $\mathbf{u}\in\{-1/\sqrt{d},1/\sqrt{d}\}^d$ with uniform distribution on the $d$-dimensional hypercube. Then he calculates a function which satisfies $M(\mathbf{y'}^d,\mathbf{u})\mathbf{y'}^d=\mathbf{u}$ and sends $M(\mathbf{y'}^d,\mathbf{u})$ to Alice. Alice applies this function to her data to obtain $\mathbf{v}=M(\mathbf{y'}^d,\mathbf{u})\mathbf{x'}^d$.

After multidimensional reconciliation step, Alice and Bob have the strings $\mathbf{v}$ and $\mathbf{u}$. Then they use the channel coding technology to correct the errors between them. Multi-edge type low density parity check (MET-LDPC) codes~\cite{richardson2002multi,richardson2008modern} are the generalization of LDPC codes, whose performances are close to Shannon's limit. Alice and Bob choose a MET-LDPC code with suitable code rate according to the parameters of quantum channel. However, the parameter estimation process has not been completed, and they do not have a prior information on the quantum channel. Fortunately, this step does not require precise results of channel characteristics. Thus, they can use the extra data to roughly estimate the parameters. These extra data are used to complete the functions of phase compensation and synchronization etc. They not only do not affect the security of CV-QKD systems, but they can also characterize the performance of quantum channel. However, because the amount of the extra data is small, the results may not be accurate. Although the error correction step does not require precise results, it will affect the performance of the information reconciliation.

Thus, we introduce another method to obtain more accurate results. For a stable CV-QKD system, the channel characteristics change very slowly. All the data of the first block are used to estimate the channel characteristics. From the second block, all the data can be used for both parameter estimation and key extraction. And the parameter estimation results of the previous block are used to assist in error correction step. The results of this method maybe more accurate because the amount of data used for parameter estimation is far larger than the former method. The parameters related to this step are mainly the signal-to-noise ratio (SNR), noise variance $\sigma^2$, quantum channel transmission $t$ etc. The first parameter is used to select a suitable MET-LDPC code. And the other parameters are used to assist in decoding. They satisfy the following relationship:
\begin{equation}
SNR=\frac{t^2V_A}{\sigma^2}.
\end{equation}

\begin{figure}
\includegraphics[width=8.8cm,height=7.0cm]{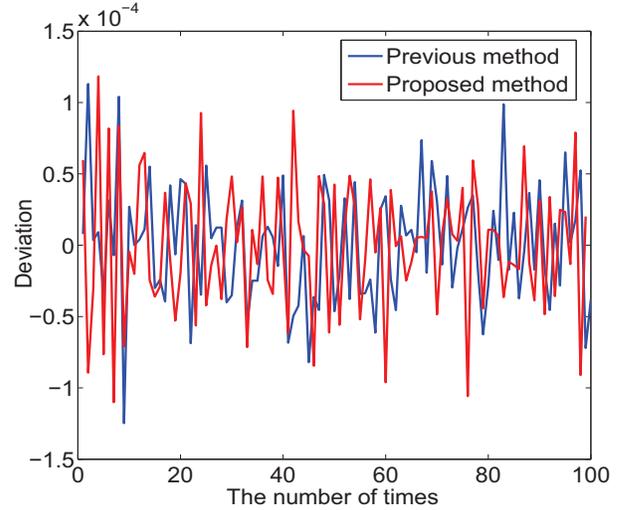}
\caption{The deviations between the estimated SNRs and the actual SNRs. The blue line refers to the results of previous method (Suppose that half of the raw data is used for parameter estimation), while the red line refers to the results of proposed method.}
\label{fig:deviation}
\end{figure}

SNR has an influence on selection of the rate of MET-LDPC codes and affects the performance of error correction. As shown in Fig~\ref{fig:deviation}, the SNRs of the two adjacent blocks of data are very close, the impact on the performance can be ignored. The deviations of the previous methods have also been shown in the Fig~\ref{fig:deviation}. The results show that the deviations are similar for both of the methods, and they are all less than $1.5\times10^{-4}$ (about 0.5\% of the actual SNRs), which has small effect on the performance of error correction. The other two parameters are used to assist decoding initialization of MET-LDPC codes. Actually, the decoding is only related to the SNRs. Therefore, as long as the SNRs are fixed, the other two parameters will not affect the decoding results. If the quantum channel is mutated, i.e. the deviations of the two adjacent blocks are quite large, then the quantum channel parameters need to be reestimated by using the first block of data after the mutation.

Reconciliation efficiency  $\beta$ refers to the error correction performance. Its value is given by $\beta=R/C$ in multidimensional reconciliation, where $R$ is the code rate of MET-LDPC and $C$ is the capacity of quantum channel, such that $C=1/2log_2(1+SNR)$. The precise $\beta$ can be obtained after parameter estimation when we get the accurate $SNR$.

Error verification step is executed to ensure the consistency of secret keys. Alice and Bob respectively compute a hash of their keys after error correction and publicly compare it. If the comparison result is the same, their keys are considered to be consistent. Otherwise, they abandon the keys. This method is worked because that even if there is only a few keys are different between them, their hash will be totally different with a high probability.

In the following, we introduce how to realize parameter estimation without publicly transmitting any raw keys when the error correction successes. In other words, the raw keys are used for both key extraction and parameter estimation. Alice (in the reverse reconciliation mode, for Bob in the direct reconciliation mode) could recover the raw keys of Bob after the error correction successes. This method is implemented based on reverse multidimensional reconciliation. Multidimensional reconciliation is used to map Gaussian distributed data to uniformly distributed binary data by rotation. Because the mapping function is reversible, the mapped binary data can be recovered to the original Gaussian data by reverse rotation. The data recovery process is as follows:

(1) Alice and Bob have the same bit strings $\mathbf{u}$ after information reconciliation. Alice computes the inverse function $M(\mathbf{y'}^d,\mathbf{u})^{-1}$ of mapping function $M(\mathbf{y'}^d,\mathbf{u})$.

(2) Alice calculates the results of the inverse function multiplied by the bit strings. Then Alice gets the normalized results of the raw keys of Bob.
\begin{equation}
\mathbf{y'}^d=M(\mathbf{y'}^d,\mathbf{u})^{-1}\mathbf{u}.
\end{equation}

(3) Alice calculates the results of $\mathbf{y'}^d$ multiplied by the norm of $\mathbf{y}^d$. It is given by
\begin{equation}
\mathbf{y}^d=\mathbf{y'}^d||\mathbf{y}^d||,
\end{equation}
where $||\mathbf{y}^d||$ is sent from Bob to Alice when Bob sends the mapping function $\mathbf{y'}^d$. All the $d$-dimensional strings $\mathbf{y}^d$ make up $\mathbf{y}$.

The security of multidimensional reconciliation has been proved. The mapping function $\mathbf{y'}^d$ does not give any information about $\mathbf{u}$, and the norm of $\mathbf{y}^d$ does not affect the security. The data recovery process does not require communications, thus it has no effect on the security of CV-QKD system.

Now, we consider the situation where error correction fails. If the error correction fails, including failing to pass error verification, Alice can not recover the raw keys of Bob. In the previous systems, the data will be abandon. However, these data can also characterize the characteristics of the quantum channel. Therefore, in order to improve the data utilization, Bob sends these data to Alice for parameter estimation. More importantly, this helps Alice and Bob to accurately grasp the characteristics of quantum channel.

Through the above process, Alice has both of the raw keys $\mathbf{x}$ and $\mathbf{y}$ whether the error correction successes or not. Then she can use the whole raw keys for parameter estimation. In other words, no raw key is wasted and it improves the data utilization.

The main purpose of parameter estimation is to estimate the secret key rate by the characteristics of quantum channel. This can be completed by the covariance matrix of the state shared by Alice and Bob. For a Gaussian modulation CV-QKD system, the covariance matrix can be given by
\begin{equation}
\Gamma=\left(
  \begin{array}{cc}
    (V_A+1)\mathbb{I}_2 & \sqrt{\eta T}Z\sigma_z \\
    \sqrt{\eta T}Z\sigma_z & (\eta TV_A+1+\eta T\xi+v_{el})\mathbb{I}_2 \\
  \end{array}
\right),
\label{eq:COM}
\end{equation}
where $V_A$ is the modulation variance of Alice, $\eta$ is the detector efficiency, $T$ is the quantum channel transmission, $\xi$ is excess noise, $v_{el}$ is electrical noise, $\mathbb{I}_2$ is a two-dimensional unit matrix, $\sigma_z$=diag(1,-1), and $Z=\sqrt{V_A^2+2V_A}$. As mentioned before, for a linear AWGN channel, the data of Alice and Bob satisfies the following relation:
\begin{equation}
\mathbf{y}=t\mathbf{x}+\mathbf{z},
\end{equation}
where $t=\sqrt{\eta T}$, and the variance of $\mathbf{z}$ is $\sigma^2=1+\eta T\xi+v_{el}$. Therefore, as shown in Eq.~\ref{eq:COM}, one only needs to estimate $t$ and $\sigma^2$. Maximum-likelihood estimators $\hat{t}$ and $\hat{\sigma}^2$ are known for the normal linear model~\cite{leverrier2010finite}:
\begin{equation}
\hat{t}=\frac{\sum_{i=1}^Nx_iy_i}{\sum_{i=1}^Nx_i^2},
\end{equation}
\begin{equation}
\hat{\sigma}^2=\frac{1}{N}\sum_{i=1}^N(y_i-\hat{t}x_i)^2,
\end{equation}
where $x_i$ and $y_i$ are $N$ (for homodyne detector, $2N$ for heterodyne detector) couples of correlated sampling Gaussian variables. The distributions of $\hat{t}$ and $\hat{\sigma}^2$ are:
\begin{equation}
\hat{t}\sim\mathcal{N}(t,\frac{\sigma^2}{\sum_{i=1}^Nx_i^2}),
\end{equation}
\begin{equation}
\frac{N\hat{\sigma}^2}{\sigma^2}\sim\chi^2(N-1),
\end{equation}
where $t$ and $\sigma^2$ are the true values of the parameters. To ensure the system security, a minimum secret key rate is obtained with a probability of 1-$\epsilon_{PE}$ when $t$ is minimum and $\sigma^2$ is maximum. In the limit of large $N$, one can get $t_{min}$ and $\sigma^2_{max}$:
\begin{equation}
\begin{array}{ll}
t_{min}\approx\hat{t}-\triangle t
\hspace{1em}
\mbox{and}
\hspace{1em}
\sigma^2_{max}\approx\hat{\sigma}^2+\triangle \sigma^2,
\end{array}
\end{equation}
where $\triangle t=z_{\epsilon_{PE}/2}\sqrt{\frac{\hat{\sigma}^2}{NV_A}}$, $\triangle \sigma^2=z_{\epsilon_{PE}/2}\frac{\hat{\sigma}^2\sqrt{2}}{\sqrt{N}}$, $z_{\epsilon_{PE}/2}$ is such that $1-\mbox{erf}(z_{\epsilon_{PE}/2}/\sqrt{2})/2=\epsilon_{PE}/2$, and the error function erf($x$) is defined as:
\begin{equation}
\mbox{erf}(x)=\frac{2}{\sqrt{\pi}}\int_0^xe^{-t^2}dt.
\end{equation}

Once the values of $t_{min}$ and $\sigma^2_{max}$ are determined, one can compute the maximum value of Holevo information $S_{\epsilon_{PE}}(y:E)$ between Bob and Eve except with a probability of $\epsilon_{PE}$. Finally, in our proposed method, when considering the finite-size effect, the secret key rate can be modified as:
\begin{equation}
K_{finite}=\beta I(x:y)-S_{\epsilon_{PE}}(y:E)-\Delta(n).
\end{equation}

Since the whole raw keys are used for both key extraction and parameter estimation, both of the secret key rate and the accuracy of parameter estimation are improved.

The last step of postprocessing is privacy amplification which is used to extract secure keys from Alice and Bob's data and eliminate the information eavesdropped by Eve. Alice randomly chooses a hash function from universal hash families and sends to Bob. Both of them use this hash function to compress their keys after error correction according to the results of parameter estimation.

\section{Performance and discussion}

We have explained how to improve the secret key rate and the accuracy of parameter estimation of CV-QKD systems. In this section, we analyse the performance of the proposed high efficiency postprocessing method for CV-QKD systems and discuss the results. For a quantum channel with unknown prior characteristics, Alice and Bob have to estimate the parameters by using the correlated data shared by them. When considering the finite-size effect, the length of the data used for parameter estimation affects the secret key rate and transmission distance. As shown in Eq.~\ref{eq:kp}, the previous methods have to sacrifice part of the raw keys, this leads to a reduction in the secret key rate and the results of parameter estimation are inaccurate.

\begin{figure}
\includegraphics[width=9.2cm,height=6.8cm]{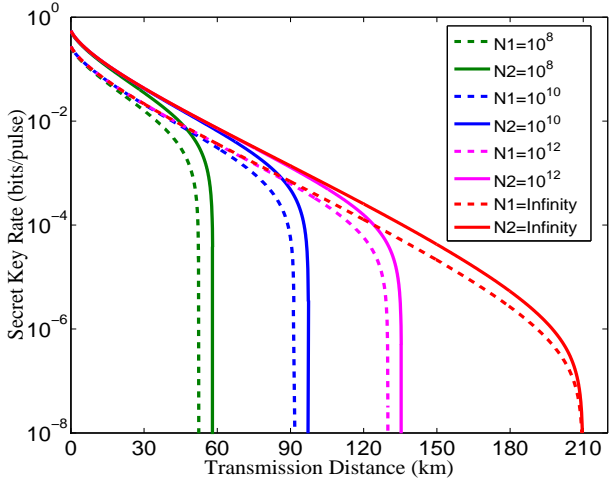}
\caption{The performance comparison between the proposed method and the previous methods. From left to right, the curves respectively represent the block lengthes with $10^8$, $10^{10}$, $10^{12}$, and the asymptotic regime. The dotted lines (N1) and the full lines (N2) represent the results the proposed method and the previous methods, respectively. The parameters of our CV-QKD system are as follows: $\xi=0.04$, $v_{el}=0.15$, $\eta=0.612$, $\beta=95\%$, and the channel loss is 0.2dB/km.}
\label{fig:results}
\end{figure}

Assume that the previous methods sacrifice half of the raw keys for parameter estimation, the other half is used for key extraction. Fig.~2 shows the performance comparison between the proposed method and the previous methods. As shown in Fig.~2, compared to the previous methods, the proposed method has higher secret key rate and longer transmission distance at different block lengthes. The main reason is that the worst case needs to be considered when estimating the parameters to ensure the security of the system. Therefore, $t_{min}$ and $\sigma^2_{max}$ are used to compute the secret key rate. However, because all the raw keys are used in the proposed method, $\triangle t$ and $\triangle \sigma^2$ are smaller than the previous methods so that a more tightness secret key rate is obtained. The results show that the secret key rate and the transmission distance are improved in the case of finite-size regime. In the asymptotic regime, half of the raw keys can also accurately estimate the parameters, thus the transmission distance is the same. While it still sacrifice half of the data, the secret key rate is also reduced by half. Thus, the high efficiency postprocessing method has a good performance when there is no prior information on the quantum channel.

\section{Conclusion}

We presented a high efficiency postprocessing method for CV-QKD, which can use all the raw keys for both parameter estimation and key extraction. This method is implemented by exchanging the execution order of parameter estimation and information reconciliation. If the error correction successes, Alice can recover the raw keys of Bob. While if the error correction fails, Bob discloses his raw keys. Thus, Alice has the raw keys of both side, she can estimate the characteristics of quantum channel without wasting any raw keys. The results show that our method improves the secret key rate and transmission distance. This method has been applied in the longest field test of CV-QKD system~\cite{zhang2017continuous}.


\begin{thebibliography}{1}

\bibitem{gisin2002quantum} N.~Gisin, G.~Ribordy, W.~Tittel, and H.~Zbinden, ``Quantum
    cryptography,'' \emph{Rev. Mod. Phys.} vol.~74, no.~1, pp. 145--195,
    2002.

\bibitem{bennet1984quantum} C. H. Bennett, and G. Brassard, ``Quantum cryptography:
    Public key distribution and cointossing,'' in \emph{Proceedings of IEEE
    Conference on Computer System and Signal Processing,} 1984, pp.
    175-179.

\bibitem{ekert1991quantum} A. K. Ekert, ``Quantum cryptography based on Bell's theorem,''
    \emph{Phys. Rev. Lett.} vol.~67, no.~6, pp. 661--663, 1991.

\bibitem{grosshans2002continuous} F. Grosshans and P. Grangier, ``Continuous variable quantum
    cryptography using coherent states,'' \emph{Phys. Rev. Lett.} vol.~88,
    no.~5, 057902, 2002.

\bibitem{weedbrook2004quantum} C. Weedbrook, A. M. Lance, W. P. Bowen, T. Symul, T. C.
    Ralph, and P. K. Lam, ``Quantum cryptography without switching,''
    \emph{Phys. Rev. Lett.} vol.~93, no.~17, 170504, 2004.

\bibitem{li2014continuous}Z. Li, Y. C. Zhang, F. Xu, X. Peng, and H. Guo,
    ``Continuous-variable measurement-device-independent quantum key
    distribution,'' \emph{Phys. Rev. A,} vol.~89, no.~5, 052301, 2014.

\bibitem{pirandola2015high}S. Pirandola, C. Ottaviani, G. Spedalieri, C. Weedbrook,
    S. L. Braunstein, S. Lloyd,  T. Gehring, C. S. Jacobsen, and U. L. Andersen, ``High-rate
    measurement-device-independent quantum cryptography,'' \emph{Nat. Photonics}
    vol.~9, pp.397-402, 2015.

\bibitem{zhang2014continuous}Y. C. Zhang, Z. Li, S. Yu, W. Gu, X. Peng, and H. Guo,
    ``Continuous-variable measurement-device-independent quantum key
    distribution using squeezed states,'' \emph{Phys. Rev. A,} vol.~90, no.~5, 052325, 2014.

\bibitem{weedbrook2012gaussian}C. Weedbrook, S. Pirandola, R. Garcia-Patron, N. J. Cerf,
    T. C. Ralph, J. H. Shapiro, and S. Lloyd, ``Gaussian quantum
    information,'' \emph{Rev. Mod. Phys.} vol.~84, no.~2, pp. 621-669, 2012.

\bibitem{jouguet2013experimental}P. Jouguet, S. Kunz-Jacques, A. Leverrier, P. Grangier, and
    E. Diamanti, ``Experimental demonstration of long-distance
    continuous-variable quantum key distribution'' \emph{Nat. Photonics}
    vol.~7, pp.378-381, 2013.

\bibitem{diamanti2015distributing}E. Diamanti, and A. Leverrier, ``Distributing secret keys
    with quantum continuous variables: principle, security and
    implementations,'' \emph{Entropy}, vol.~17, no.~9, pp. 6072-6092, 2015.

\bibitem{zhang2017continuous}Y. C. Zhang, Z. Li, Z. Chen, C. Weedbrook, Y. Zhao, X.
    Wang, C. Xu, X. Zhang, Z. Wang, M. Li, X. Zhang, Z. Zheng, B. Chu, X.
    Gao, N. Meng, W. Cai, Z. Wang, G. Wang, S. Yu, and H. Guo,
    ``Continuous-variable QKD over 50km commercial fiber,'' Preprint at arXiv:1709.04618, 2017.

\bibitem{WangHSEC}X. Wang, Y. C. Zhang, S. Yu, and H. Guo,
    ``High speed error correction for continuous-variable quantum key distribution with
    multi-edge type LDPC code,'' Preprint at arXiv:1711.01783, 2017.

\bibitem{Yang2017FPGA}S. S. Yang, Z. L. Bai, X. Y. Wang, and Y. M. Li, ``FPGA-Based
    Implementation of Size-Adaptive Privacy Amplification
    in Quantum Key Distribution,'' \emph{IEEE Photonics Journal,} vol.~9, no.~6, pp. 1-8, 2017.

\bibitem{Wang2018PA}X. Wang, Y. C. Zhang, S. Yu, and H. Guo, ``High-speed Implementation of
    Length-compatible Privacy Amplification in Continuous-variable Quantum Key Distribution,'' \emph{IEEE Photonics Journal,}
    vol.~10, no.~3, 2018.

\bibitem{leverrier2010finite} A. Leverrier, F. Grosshans and P. Grangier, ``
    Finite-size analysis of a continuous-variable quantum key distribution,''
    \emph{Phys. Rev. A,} vol.~81, no.~6, 062343, 2010.

\bibitem{jouguet2012analysis} P. Jouguet, S. Kunz-Jacques, E. Diamanti and A. Leverrier,
    ``Analysis of imperfections in practical continuous-variable
    quantum key distribution,'' \emph{Phys. Rev. A,} vol.~86, no.~3, 032309, 2012.

\bibitem{zhang2017finite} X. Zhang, Y. Zhang, Y. Zhao, X. Wang, S. Yu, and
    H. Guo, ``Finite-size analysis of continuous-variable
    measurement-device-independent quantum key distribution,'' \emph{Physical Review
    A,} vol.~96, no.~4, 042334, 2017.

\bibitem{papanastasiou2017finite} P. Papanastasiou, C. Ottaviani, and S. Pirandola, ``Finite-size analysis
    of measurement-device-independent quantum cryptography with continuous variables,''
    \emph{Physical Review A,} vol.~96, no.~4, 042332, 2017.

\bibitem{lupo2017parameter} C. Lupo, C. Ottaviani, P. Papanastasiou, and S. Pirandola, ``Parameter estimation
    with almost no public communication for continuous-variable quantum key distribution,'' Preprint at arXiv:1712.00743, 2017.

\bibitem{leverrier2015composable} A. Leverrier, ``Composable security proof for continuous-variable quantum key distribution with coherent states,''
     \emph{Phys. Rev. Lett.} vol.~114, no.~7, 070501, 2015.

\bibitem{van2004reconciliation} G. V. Assche, J. Cardinal and N. J. Cerf, ``
    Reconciliation of a quantum-distributed Gaussian key,'' \emph{IEEE Transactions
     on Information Theory,} vol.~50, no.~2, pp. 394-400, 2004.

\bibitem{wang2017efficient} X. Wang, Y. C. Zhang, Z. Li, B. Xu, S. Yu, and H. Guo,
    ¡°Efficient rate-adaptive reconciliation for continuous-variable quantum
    key distribution,¡± \emph{Quantum Inf. Comput.} vol.~17, no.~13\&14, pp. 1123-1134, 2017.

\bibitem{leverrier2008multidimensional} A. Leverrier, R. All\'{e}aume, J. Boutros, G. Z\'{e}mor, and P.
    Grangier, ``Multidimensional reconciliation for a continuous-variable
    quantum key distribution,'' \emph{Phys. Rev. A,} vol. 77, no.~4, 042325
    2008.

\bibitem{bennett1988privacy}C. H. Bennett, G. Brassard, and J. M. Robert,
    ``Privacy amplification by public discussion,'' \emph{SIAM journal on
    Computing,} vol.~17, no.~2, pp. 210-229, 1988.

\bibitem{bennett1995generalized} C. H. Bennett, G. Brassard, C. Cr\'{e}peau and U. M. Maurer,
    ``Generalized privacy amplification,'' \emph{IEEE
    Transactions on Information Theory,} vol.~41, no.~6, pp. 1915-1923, 1995.

\bibitem{gallager1962low} R. Gallager, ``Low-density parity-check codes,'' \emph{IRE Transactions on information theory,} vol.~8, no.~1, pp. 21-28, 1962.

\bibitem{richardson2008modern} T. Richardson, and R. Urbanke, ``Modern coding theory,'' Cambridge university press (New York), 2008.

\bibitem{grosshans2003quantum} F. Grosshans, G. Van Assche, J.Wenger, R. Brouri, N. J.
Cerf, and P. Grangier, ``Quantum key distribution using gaussian-modulated coherent states,'' \emph{Nat.,} vol.~421, pp. 238-241, 2003.

\bibitem{richardson2002multi} T. Richardson and R. Urbanke, ``Multi-edge type LDPC
    codes,'' presented at \emph{Workshop honoring Prof. Bob McEliece on his 60th
    birthday,} California Institute of Technology, Pasadena, California, pp.
    24-25, 2002.

\end{thebibliography}
\end{document}